\documentclass[referee]{aa}
\usepackage{graphics}
\usepackage{psfig}

\begin{document}

\title{Astrophysical Lasers Operating in optical Fe II
                            Lines
                in Stellar Ejecta of $\eta$ Carinae}

\author{ S. Johansson \inst{1}
  \and V.S. Letokhov \inst{2,1}}

\offprints{S. Johansson}

\institute{Lund Observatory, Lund University, 
            P.O. Box 43, S-22100 Lund, Sweden \\
            email: sveneric.johansson@fysik.lu.se,
                   Vladilen.Letokhov@fysik.lu.se
   \and Institute of Spectroscopy, Russian Academy of Sciences, Troitsk, 
     Moscow region, 142190, Russia}

\date{Received <date> / Accepted <date>}

\titlerunning{Astrophysical Lasers in $\eta$ Carinae}
\authorrunning{Johansson \& Letokhov}

\abstract{After the discovery of space masers based on
OH  radicals (Weaver et al., 1965) and H$_2$O (Cheung  et  al.,
1969) such microwave lasers have been found to work in  more
than 100 molecular species (Elitzur, 1992; Townes, 1997), as
well  as  in  highly  excited H atoms (Strelnitski  et  al.,
1996).  In  the  IR region (10 $\mu$m), the effect of  stimulated
emission   of  radiation  in  the  CO$_2$  molecule  has   been
discovered in the Martian and Venus' atmospheres  (Betz  et
al.,  1976;  Mumma  et al., 1981). We  report  here  on  the
discovery  of laser action in the range 0.9-2 $\mu$m in  several
spectral   lines  of  Fe  II,  which  are  associated   with
transitions  from  ``pseudo-metastable'' states  populated  by
spontaneous  transitions from Ly$\alpha$ pumped Fe  II  levels.  The
intense  Ly$\alpha$  radiation is formed in the HII  region  of  gas
condensations close to the star $\eta$ Car. The laser  transitions
form  together with spontaneous transitions closed radiative
cycles, one of which includes the extremely bright 2507/09 \AA\
lines.  Closed radiative cycles, together with an accidental  mixing  of
energy  levels, may provide an explanation of  the  abnormal
intensities  of  these  UV  non-lasing  lines.   
Using   the complicated  energy level diagram of Fe II we present  those
peculiar  features,  which are essential  for  the  inverted
population and laser effect: the pumping, the level  mixing,
and  the  ``bottle  neck'' for spontaneous  decay.  The  laser
action  is  a new indicator of non-equilibrium and spatially
non-homogeneous  physical  conditions  as  well  as  a  high
brightness temperature of Ly$\alpha$ in ejecta from eruptive  stars.
Such  conditions  are very difficult to  probe  by  existing
methods, and we propose some future experiments. The fact, that the 
lasing near-IR lines appear in the spectrum with about the same intensity as
non-lasing lines is discussed and compared with the situation in masers. 
}

\maketitle

\section{Introduction}

A  gas  condensation  formed by  expelled  material  in  the
neighborhood  of  a  bright star  is  an  optimal  site  for
hydrogen  atoms  to  be photoionized by extreme  ultraviolet
(EUV)  radiation under high-vacuum conditions.  The  central
star  emits  the EUV radiation, and hydrogen is  ionized  by
photons with $\lambda <$ 912 \AA. Collisions with electrons at a low rate
yield  the  characteristic recombination spectrum of  atomic
hydrogen.  The  intense Ly$\alpha$ line at 1215 \AA, which  cannot  be
observed  from  Earth  because of  interstellar  absorption,
contributes  about  70 \%  of the  energy  radiated  by  the
recombination  lines.  However, Ly$\alpha$  radiation  is  observed
indirectly  through its interaction with atoms and  ions  of
other     elements,     particularly    through     resonant
photoexcitation  due  to wavelength  coincidences.  Such  an
accidental  coincidence of spectral lines (the Bowen  (1935)
mechanism) is a rare phenomenon. Since the spectral lines of
free atoms and ions are very narrow the Bowen mechanism  has
no  practical  use  in  present-day  lasers.  However,  this
selective excitation mechanism generates some spectral lines
in  astrophysical plasmas, which contain a mixture of  atoms
and  ions  exposed  to intense spectral lines  of  hydrogen,
helium and other abundant light elements. The possibility of
laser action in stellar atmospheres based on processes  such
as  the  Bowen  mechanism  was considered  fairly  long  ago
(Letokhov,  1972;  Lavrinovich \& Letokhov,  1974),  but  the
resulting  line emission has been difficult to resolve  from
the stellar background radiation.
%
\\
\\
\noindent \small 
     Fig. 1. Illustration of the main radiative processes in the
     blobs  outside $\eta$ Car, involving photoionization  of  HI
     and  Fe I by stellar black-body radiation and subsequent
     resonant excitation of Fe II by intense H Ly$\alpha$  radiation.
\normalsize
\\

     The observational conditions changed radically with the
launch  of  the {\em Hubble Space Telescope (HST)}  and  the  High
Resolution  Spectrograph  (HRS). Its  successor,  the  Space
Telescope   Imaging  Spectrograph  (STIS),   provides   high
spectral  resolution in a broad wavelength range (1150-10400
\AA)  and a near-diffraction-limited angular resolution  ($\sim$ 0.1
arcsec)  (Kimble  et  al., 1998).  Beginning  in  1998,  the
eruptive  star $\eta$ Car  became one of  the  most  remarkable
targets  for this instrument (Gull et al., 1999). It  is  an
exceptionally  interesting  astrophysical  object  expelling
enormous amounts of material into its surroundings.  In  the
immediate  vicinity of the central star  at  a  distance  of
100-1000   stellar   radii  ($r_{\mathrm s}=3\cdot 10^{13}$  cm),
   compact   gas
condensations   (GC), called the Weigelt blobs B, C and  D,
of   exceptional brightness  have  been  discovered 
(Weigelt  \&  Ebersberger, 1986).  They
are  unique  astrophysical plasmas having a higher  hydrogen
density  ($N_{\mathrm H}\approx 10^7-10^{10}$ cm$^{-3}$) (Davidson  \&  Humphreys,
1997) than planetary nebulae ($N_{\mathrm H}\approx 10^4$ cm$^{-3}$) and located only
a few light days away from the central star.

      For  clarity we illustrate in Fig. 1 the  sequence  of
basic  radiative processes in the blobs, initiated  by  stellar
blackbody  (Planck)  radiation acting  on  H I  and  Fe I  and
followed  by  pumping due to an accidental  resonance  (PAR)
between FeII and Ly$\alpha$. A modelling of the spectrum of the 
Weigelt blobs (Verner et al 2002) has verified the importance of Ly$\alpha$ 
pumping to reproduce the fluorescent Fe II lines and continuum pumping to 
reproduce low-excitation Fe II as well as [Fe II] lines. 
As $N_{\mathrm H}$ is relatively high, the blob will
almost completely absorb the Lyman-continuum radiation  from
the  central  star.  The  Str\"{o}mgren border,  separating  the
regions  of  completely  ionized (HII  region)  and  neutral
hydrogen  (HI region), is thus located inside the blob.  The
HII  region  is adjacent to the HI region. The intensity  of
the Ly$\alpha$ recombination line in the HII region of the blob  is
10$^3-10^4$ times higher than that coming from the central  star
(Klimov et al., 2002; Johansson and Letokhov, 2001b). We get
a  unique  situation  in  the HI region  where  a  partially
ionized  mixture of many elements is exposed to  an  intense
Ly$\alpha$ flux coming from the HII region. This is
similar  to  a typical ``laser'' situation where  the  pumping
flash lamp irradiates the nearby active laser medium.

Preliminary data on {\em laser effects in the Weigelt blobs} 
 have been published in the form of short
letters  (Johansson  \& Letokhov, 2002,  2003).  The  present
paper  contains  a  more  detailed treatment  of  the  laser
schemes  related to the energy level diagram  of  Fe  II,  a
detailed  analysis of the amplification in  various  schemes
and a discussion of the spectral width of laser lines.  
Below we give a brief outline of the paper.

The iron atoms occur in the HI region at a density
of  about $N_{\mathrm {Fe}} \approx 10^{-4}\cdot N_{\mathrm H}$, and 
they are easily photo-ionized by the
stellar  photospheric radiation in the spectral  window  7.9
eV$<$h$\nu<$13.6 eV (Fig. 1). The Fe$^+$ ion has a line-rich spectrum,
and some lines coincide in wavelength with the wide Ly$\alpha$ line
resulting  in  a selective photo-excitation to  high  energy
levels  (Brown et al.,  1979;  Johansson  \&
Jordan, 1984). This is discussed in Sec. 2 as well as a general scheme of
the radiative decay routes of the Ly$\alpha$-excited Fe  II  levels. 
These radiative pathways are  numerous and peculiar  because  of  the
complex  Fe II energy level diagram, which offers both  {\em accidental
level  mixing  (ALM)} and the presence of ``{\em pseudo-metastable}''
$(PM)$  high-lying  states  (Sec. 3).  The  latter   have
lifetimes  of about 1 $\mu$s -1 ms and they create ``bottle-necks''
in  the  chain of spontaneous radiative decays of  the Ly$\alpha$-excited Fe II levels.

In Sec. 4 the formation of population inversion in the PM states of 
Fe II, a basic requirement for lasing effect, is described as well as the size of the 
amplification coefficient. Two types of ``bottle-neck'', occurring in PM states 
with ms and $\mu$s lifetimes, respectively, are discussed separately in great 
detail in Secs. 5 and 6. The ms-``bottle-neck'' (Sec. 5) provides a) a strong 
radiative cycle in Fe II generated by Ly$\alpha$ pumping, b) lasing in spectral 
lines at 1.68 $\mu$m and 1.74 $\mu$m, and c) the basis for the explanation of 
the anomalously bright non-lasing UV lines at 2507/09 \AA. The $\mu$s ``bottle-neck''
(Sec. 6) provides laser effect in several lines in the range 0.9-1.0 $\mu$m,
observed in {\em HST/STIS} spectra of the Weigelt blobs. For completeness, we are
also considering (Sec. 7) the combination of strong and weak radiative cycles,
where a ms-``bottle-neck'' creates strong lasing lines at $\lambda\lambda$9617/9913.
For the $\lambda\lambda$ 9617/9391 lines we also observe a peculiar intensity ratio, which is in disagreement with the predicted branching ratio.
In Sec. 8 we briefly discuss potential sources of Ly$\alpha$ radiation providing
the pumping of the Fe$^+$ ions in the Weigelt blobs.

Sections 9 and 10 contain a qualitative discussion of the differences
between astrophysical lasers and masers from an observational point of view. The
evidence for astrophysical masers is mostly based on their high brightness and narrow
line widths, observed by high-resolution radio telescopes. However, the astrophysical
laser lines may not brighter than ordinary lines, and the high spectral resolution
($\simeq 10^5$) of optical telescopes/spectrometers (e.g. {\em HST/STIS}) is too small
to observe their narrow Doppler profiles. The possibility of using laser heterodyne
Brown-Twiss-Townes interferometry is proposed in Sec. 10.

\section{Radiative excitation and relaxation of Fe II levels populated
by absorbed Ly$\alpha$ radiation.} 

A  large  number of observed as well as predicted absorption
lines of Fe II fall within a spectral width of 3\AA\ around Ly$\alpha$.
The  transitions start from low metastable states as
illustrated in Fig. 2 (see Johansson \& Jordan, 1984). There are about  15
emission lines/\AA\ observed in the laboratory spectrum  of  Fe
II  in  the  wavelength  region around  Ly$\alpha$.  The  number  is
probably  even  higher for absorption lines, some  of  which
could   be   pumped   by   Ly$\alpha$    in  astrophysical   plasmas.
Consequently, numerous Ly$\alpha$  pumped fluorescence lines of Fe II
have  been  identified in spectra of various  emission  line
sources,  eg.  chromospheres  of  the  sun  and  cool  stars
(Jordan,  1988a,  b; Harper et al., 2001),  symbiotic  stars
(Hartman  \&  Johansson 2000), and the environment  of   AGNs
(Netzer, 1988).

The  Weigelt blobs outside $\eta$ Car are rather special because
of  their  high  optical density for Ly$\alpha$ ,
$\tau_{\alpha} \approx 10^8$, and  their  short
distance to the central star. The
ratio  of  the Einstein coefficients for Ly$\alpha$  and  the  pumped
absorption  lines of Fe II is  $\approx 10^2$, and since all Fe atoms 
in  the  HI region are ionized the abundance ratio 
$N_{\mathrm {Fe^+}}/N_{\mathrm H}\approx 10^{-4}$.  
Since the optical density for the Fe II absorption lines is in the
range $\tau_{\mathrm {Fe II}} \approx $ 10 -100,  
the  resonant intervals  within  the  wide  Ly$\alpha$ 
profile  are  fully  absorbed.  A  significant  density   of
absorption lines should result in a total absorption  of  Ly$\alpha$ 
in the H I region of the blob. 

 The  photoselective excitation  rate  
of  state  {\it 4}  is defined by
\begin{equation}
W^{14}_{\mathrm {exc}} = A_{41}\frac{g_4}{g_1}\left[\mathrm{exp}\left(\frac{h\nu_{14}}
{kT_{\alpha}}\right)-1\right]^{-1},
\end{equation}
where  the indices 1 and 4 correspond to the level notations
in  Figs.  2b,c.  The  Einstein  coefficient  for
spontaneous  decay  is $A_{41} = 1.2\cdot10^7$ s$^{-1}$ 
(Kurucz  2003),  and
$T_{\alpha}=T$(Ly$\alpha$) is the brightness (or effective) 
temperature  of  Ly$\alpha$ 
inside  blob  B. According to the qualitative picture  given
above  the  stellar Lyman continuum radiation is transformed
into intense Ly$\alpha$ radiation. A more detailed analysis of this
{\em ``spectral  compression'' effect} is presented in (Johansson  \&
Letokhov, 2004), and it shows that $T_{\alpha}$ {\em is comparable with  the
effective  temperature of the stellar  photosphere}.  In  the
case  of  $\eta$ Car, we adopt k$T_{\alpha} \approx 1.0-1.5$ eV
($T_{\alpha}\approx (12-18)\cdot 10^3$ K),
which  means a photoselective excitation rate of 
$W^{14}_{\mathrm {exc}}\approx 10^3 - 10^4$ s$^{-1}$.  Let  us
emphasize that without compensation for the dilution  factor
by  the ``spectral compression'' effect the value of $W^{14}_{\mathrm {exc}}$ will
be  about  1s$^{-1}$, making it impossible to provide an inverted
population in the 3$\rightarrow$2 transition.
%
\\
\\
\noindent \small
Fig. 2. Three types of radiative decay schemes of high-lying
     Fe  II levels photoexcited by Ly$\alpha$: {\bf a}) a fast radiative
     decay  down to metastable states via short-lived levels
     without  any  ``bottle  neck'' effect;  {\bf b})  a  radiative
     decay  via high, pseudo-metastable states (the  ``bottle
     neck''   effect)   with   formation   of   an   inverted
     population;  {\bf c}) a radiative decay with  ``bottle  neck''
     effect,  inversion  of population  and  return  to  the
     initial state.
%
\\
\\
\normalsize
      The   numerous  possible  pathways  of  the  radiative
relaxation  (decay) of the Ly$\alpha$-excited Fe II levels  can  be
grouped  in two categories as illustrated in Fig. 2:
{\bf
\begin{itemize}
\item  a spontaneous decay back to the low metastable states  without
any accumulation in higher states (Fig. 2a); 
\item spontaneous
decay  via  long-lived higher states (``bottle neck''),  which
cause  an  inverted population with an associated stimulated
emission (Fig. 2b,c). 
\end{itemize}}
The first type of pathway (Fig. 2a) is
dominating  for  all  photoexcited  states,  and  the   type
illustrated in Fig. 2b occurs as a result of $ALM$,  discussed
in  section 2. The general case is a combination of (1)  and
(2),  where  the branching fractions from level {\it 4}  determine
the  population in {\it 3}. The branching fractions for the  decay
feeding  the  long-lived states are thus determined  by  the
strength  of  the level mixing.  In this paper we  focus  on
case  (2),  i.e. those Ly$\alpha$ pumped Fe II levels  having  large
branching fractions to feed the $PM$ states. It should be pointed out
that the continuum pumping, shown by Verner et al (2002) to increase the
population of the low metastable states through cascading, cannot directly
 populate the PM states due to parity considerations.

         If level {\it 2} in case (2) decays to the initial state,
i.e.  the  level being pumped, we obtain a closed  radiative
cycle  (Johansson \& Letokhov, 2003). This is  determined  by
the  possible decay channels of state {\it 2} and their  branching
fractions,  which may vary among the set of  short-lived  
states  (lifetime of a few ns) fed by the decay from the  $PM$
states.  The distribution of the decay of state  {\it 2}  is  also
important  for permanent pumping of the low-lying metastable
states,  with  lifetimes  of the  order  of  one  second.  Two
examples  of  pathways containing radiative ``bottle  necks'',
according  to  the  scheme in Fig. 2b,c,  are  discussed  in
detail  in Sections 5-6 and a combination of them in  Sec. 7.
They  represent closed radiative cycles, but with  different
efficiencies, as the branching fractions differ considerably
in  those  transitions, that bring them back to the  initial
level.  These  two  pathways of radiative  decay  have  also
different lifetimes of the ``bottle neck'' levels 3' and  3":  a
ms-lifetime for level {\it 3'} and a $\mu$s-lifetime for {\it 3"} (Fig.  2b,c).
The  branching fractions for the weakly closed cycle  (0.1\%)
and  for  the  strongly closed cycle (85\%) are indicated  in
Fig. 2b,c.
%
\\
\\
\noindent \small
Fig. 3. Time   scale   of  spontaneous   radiative   and
     collisional relaxation of highly excited Fe II  states.
     An   intermediate   range  including  pseudo-metastable
     states  is  favourable obtaining an inverted population
     in the 3$\rightarrow$2 transition.
\\
\\
\normalsize
 The  rate  of  collisions between excited Fe$^+$  ions  and  H
atoms, with a maximum density of 10$^{10}$ cm$^{-3}$, is less than one
per  second. The presence of forbidden lines from the  long-
lived metastable states (lifetime of about 1s) verifies that
the  relaxation of the excited Fe$^+$ ions is purely  radiative
(Fig. 3). Most important are the ``closed'' radiative pathways
including  Ly$\alpha$   pumping, spontaneous decays  and  stimulated
emission,  since  they  provide a  collision-free  and  pure
radiative  conversion  of intense  Ly$\alpha$  radiation  into  the
intense  and anomalous $\lambda\lambda$2507/09 UV lines of Fe II  (Johansson
\& Letokhov, 2003).

\section{Spectral peculiarities in Fe II.}

           Two properties, one atomic and one astrophysical,
make  the  spectrum  of Fe II very dominant  in  spectra  of
cosmic  sources.  A complex atomic structure  with  a  large
density  of energy levels and the high cosmic iron abundance
result  in numerous lines in ultraviolet and optical spectra
- absorption lines in stellar atmospheres and emission lines
in  nebular  regions.   Although Fe  II  has  been  thoroughly
studied  in the laboratory and more than 1000 energy  levels
have  been  found, ultraviolet stellar spectra still  contain
unidentified lines of Fe II.

     The  specific atomic astrophysics problem discussed  in
the  present  paper involves two special features  that  are
extremely  rare  in  simple spectra but  characteristic  for
complex  spectra:  {\em accidental level  mixing  (ALM)}  and  the
presence of ``{\em pseudo-metastable}'' $(PM)$ states. Level mixing is
a  well-known effect, which in its simplest form depends  on
deviations  from a pure coupling scheme for  the  electrons,
and  a final state is described as a linear combination  of,
for  example, pure $LS$ states. Such mixing of levels  may  be
well   predicted   by  theory  even  for  complex   systems.
Perturbation  and level mixing as a result of  configuration
interaction  are also in general taken care of  by  detailed
calculations.  However, two energy levels  having  the  same
parity  and  $J$-value may also mix if they accidentally  fall
very close to each other in energy. They can have completely
different    $LS$    character   and   belong   to    different
configurations. Such an $ALM$ is very difficult to predict  in
quantum mechanical calculations.
%
\\
\\
\noindent  \small
Fig. 4. Partial energy level diagram showing some of  the
      3d$^6nl$  configurations of Fe II, where $nl =4p, 5s,  4d,$
      and $5p$. Each configuration is represented by a number
      of  boxes  connected by a thin, dotted line.  One  box
      contains   all  (between  2  and  37)  energy   levels
      obtained when adding one electron ($nl$) to one  of  the
      3d$^6$  parent  terms in Fe III. The $4p$ and $5p$  are  odd-
      parity  configurations; the others are of even parity.
      The  lowest  dashed  box includes  the  62  metastable
      levels,  and  the  smaller  box  12  pseudo-metastable
      levels  (see  text, Sec. 3). We also  mark  the  level
      mixing  occurring between levels in the  left-most  $5p$
      box and two high $4p$ boxes, all of odd parity.
\\
\\
\normalsize
     The presence of $PM$ states is the crucial feature in the
radiative  processes discussed in this paper, but before  we
define  the  concept ``pseudo-metastable''  we  give  a  brief
description of metastable states. A metastable state  in  an
atom(ion) is a state that cannot decay to a lower  state  by
electric  dipole (E1) radiation, i.e. obeying the  selection
rules  for  parity and $J$-value. Hence, all  states  of  even
parity that are located below the lowest state of odd parity
are  by  definition  metastable,  and  they  have  radiative
lifetimes  of  the  order  of 1 s.  Metastable  states  are
generally   collisionally  deexcited  but   in   low-density
astrophysical plasmas the probability for collisions is very
low.  Under  such  nebular conditions the metastable  states
decay  radiatively in electric quadrupole (E2)  or  magnetic
dipole  (M1) transitions observed as forbidden lines.  There
are  62  metastable states in Fe~II belonging to  the  three
configurations 3d$^6$4s, 3d$^7$ and 3d$^5$4s$^2$. One of  the  three  low
even-parity configurations containing the metastable  states
is  the  ground configuration 3d$^6$4s. In Fig.  4  we  show  a
condensed energy level diagram for Fe II, arranged according
to the ``branching rule'' for the build-up of the structure of
complex spectra. Thus, by starting with the next higher ion,
Fe  III,  we  add a $4s$-electron to each $LS$-term (the  parent
term)  of  the  ground configuration 3d$^6$. The  resulting  $LS$
terms  in  Fe II are represented by one box for each  parent
term. All $4s$ boxes are connected by a dotted line in Fig. 4.
Since  the  $4s$ binding energy is independent of  the  parent
term the pattern of connected $4s$ boxes mimics the pattern of
the  parent  terms.  The next higher  configuration  in  the
diagram  is the odd-parity $4p$ configuration, represented  by
grey boxes. Thus, according to the definition all levels  in
the   boxes  located  below  the  lowest  grey  $4p$  box  are
metastable. These are placed in a marked box in the diagram.

     Due  to the extended structure of the parent terms  the
$4s$ configuration extends above $4p$, i.e. the highest $4s$ boxes
are located above the lowest $4p$ box (see Fig. 4). This is  a
special  feature of the transition elements. In  Fe  II  six
observed   (and   two  unknown)  $LS$-terms   of   the   3d$^6$4s
configuration  are  located above the  lowest  $LS$  terms  of
3d$^6$4p,  which means that a $4s$ electron can make  a  $downward$
transition  to  a $4p$ state. The lifetime of  these  high  $4s$
states  is  of the order of 1 ms, three orders of  magnitude
shorter  than for pure, metastable states but six orders  of
magnitude longer than for ordinary, excited states.  Because
of their long lifetimes we call them $PM$ states, and they are
marked  in  a  dashed box in Fig. 4. It looks  like  the  PM
states  keep a fraction of the metastability of their sister
LS terms in the same configuration, even if $LS$ -allowed decay
channels   become   available.  Without   looking   at   the
spectroscopic origin of these $PM$ states, but just  their  $LS$ 
assignment,  it  is  not possible to distinguish  them  from
other $LS$  terms. However, the transition probability of their
decay  to  4p  is six orders of magnitude smaller  than  for
normal  allowed transitions. It turns out that this  is  the
crucial  feature behind the explanation of the laser  action
in  Fe  II observed in $\eta$ Car and discussed in detail in  this
paper.

%
\noindent \small 
\\
\\   
Fig. 5. Schematic energy level diagrams illustrating the
      three  cases  of  radiative cycles discussed  in  this
      paper:  {\bf a}) a real case in Fe II with a combination  of
      the  models  illustrated in Figs  2a  and  b,  and  {\bf b})
      another Fe II case illustrated in Fig 2c.
\\
\\
\normalsize
     The  two  cases of four-level schemes in Fe II involved
in  the  radiative  processes discussed in  the  forthcoming
sections of this paper are shown in Figs. 5a and b. The  low
4s  box (at 1eV) in Fig. 5a is the a$^4$D term in the lowest 4s
box  in  Fig. 4. It can be pumped by Ly$\alpha$  to several 5p levels
around 11 eV, which is discussed in detail in Sec. 3. Due to
a near coincidence in energy there is an $ALM$ between some of
the  lowest $5p$ levels and some high $4p$ levels, as  indicated
in  Fig.  4,  and  a mixing of states from  two  consecutive
members  ($4p$ and $5p$) in a Rydberg series is very  rare.  The
$ALM$  means that the high $4p$ levels (denoted 4p" in Fig. 5a),
which  belong  to  $LS$   terms  of  the  (b$^3$F)4p  and  (b$^3$P)4p
subconfigurations, are also radiatively populated by Ly$\alpha$. The
fast decay from $5p$ is to $5s$ (in the near-IR) and back to  $4s$
via  $4p$. The fast decay from $4p$" is down to $4s$" in Fig.  5a,
i.e. along the vertical line down to (b$^3$F)4s and (b$^3$P)4s  in
Fig.   4.   As  all  other  strong  $4s-4p$  transitions   the
corresponding lines fall around 2500 \AA. These $4s"$ levels are
the  $PM$  states discussed above, and they can decay in  slow
transitions to the low $4p$ states at 5 eV (40000  cm$^{-1}$).  The
$4p$ levels have fast decays back to the metastable $4s$ levels,
as shown in Fig. 5a.

     The  case outlined by the scheme in Fig. 5b starts from
a  higher metastable $4s$ level (marked 4s'), which is excited
up  to  a $4s4p$ state by Ly$\alpha$  radiation. The natural decay  for
this pumped state is down to a 4s$^2$ state at about 7 eV. This
level  belongs  to the 3d$^5$4s$^2$ configuration,  which  is  not
included in Fig. 4. It has about the same properties as  the
$PM$  states  discussed  in  the first  case  above,  but  the
lifetime  is  about  three orders of magnitude  smaller,  or
about  1 $\mu$s. The 4s$^2$ state decays in a ``two-electron jump''  to
$4p$,  i.e. to the same $4p$ levels as for the $4s$" state in  the
first case. The $4p$ level has a branch to the metastable  $4s$'
level  and even in this case we can find a closed cycle  for
the radiative excitation and deexcitation.

\section{Population inversion and amplification coefficient in
  the ``bottle-neck'' 3$\rightarrow$2 transition.}

There  are  a  number of decay schemes in Fe II creating  an
inverted  population due to a radiative ``bottle  neck''.  The
main  schemes  are  shown in Fig. 5a,b, where  few  selected
energy levels and radiative transitions from the level-  and
line-rich spectrum of the complex system of Fe II illustrate
a  photo-selective excitation by Ly$\alpha$  and a subsequent cascade
of  radiative  decay channels. A remarkable feature  in  the
decay  schemes  is  the existence of $PM$ states.  They  decay
relatively slowly (radiative lifetimes from 10 $\mu$s to 1 ms) in
permitted  transitions  to {\em lower,  rapidly  decaying}  states
(lifetimes of a few ns). However, the time elapsing  between
successive collisions between excited Fe$^+$ ions and H  atoms,
the  main  component  of the GC with a  maximum  density  of
$N_{\mathrm H}= 10^{10}$ cm$^{-3}$,  is one second or even longer.  Therefore,  the
relaxation  of  the  Ly$\alpha$   excited Fe$^+$  ions  occurs  without
collisions in a purely radiative way.

       The  cascade  of radiative decay of the  Ly$\alpha$   excited
states  is  characterized by a peculiar  time  hierarchy  as
illustrated in Fig. 3. On the time scale given in the figure
one can distinguish between three relaxation regions: a fast
spontaneous  decay, a slow spontaneous  decay,  and  a  much
slower  collisional relaxation. Therefore,  the  spontaneous
radiative   decay  from  the  Ly$\alpha$   pumped   states   results
inevitably in an inverted population in the transition  from
the $PM$ state (the 3$\rightarrow$2 transition in Fig. 5), which acts  as
a  ``bottle-neck''  for  radiative decay  in  the  absence  of
collisions.  If  the  inverted population  density  and  the
optical  path  length for this transition are large  enough,
the   spontaneous  radiation  will  be  amplified   due   to
stimulated  emission in the atomic ensemble. The  stimulated
emission  enhances the intensity of the weak lines generated
by  the  slow spontaneous radiative decay of the $PM$  states,
and  it  becomes comparable with the intensity of the  fast,
spontaneous transitions populating them.

      A steady state population inversion is thus achieved in the
3$\rightarrow$2 transition with a density of     
\begin{equation}
\Delta N \approx N_3 - N_2 \approx N_3,
\end{equation}
since N$_2<<$N$_3$  due to the much faster decay of level {\it 2}.
The small influence of the statistical weights has
been neglected as their ratio is close to one. Since the short-lived levels 
{\it 4} and {\it 2} can be regarded as empty relative to the long-lived 
levels {\it 1} and {\it 3} the radiative excitation $W^{14}_{\mathrm {exc}}$ 
produces the following steady-state  distribution of the populations $N_1$ 
(in the initial level {\it 1}) and $N_3$ (in the long-lived excited level {\it 3}):
\begin{equation}
N_3 = N_1 W^{14}_{\mathrm {exc}}\tau_3,
\end{equation}
where $A_{43}>>W^{14}_{\mathrm {exc}}$. Since the rate of stimulated emission in the 
3$\rightarrow$2 transition, $W_{32}$, is not included in Eq. 3, all Fe II ions can be accumulated in level {\it 3} if $W^{14}_{\mathrm {exc}} >> 1/\tau_3$ ($N_3>>N_1$).  
In the lasing volume, however, $W_{32}$ can be much higher than 1$/\tau^3$, and a fast
closed radiative cycle will therefore maintain the population of level {\it 1} in such
a way that
\begin{equation}
N_3 = N_1 \frac{W^{14}_{\mathrm {exc}}}{W^{14}_{\mathrm {exc}}+\tau^{-1}_3},
\end{equation}
or even
\begin{equation}
N_3 = N_1, 
\end{equation}
when the excitation rate is large ($W^{14}_{\mathrm {exc}} >> 1/\tau_3$). We restrict ourselves to make an estimation of the initial amplification without including stimulated transitions (lasing), which will be the subject of future work.

The linear amplification coefficient for the  3$\rightarrow$2
transition is defined by the standard expression
\begin{equation}
\alpha_{32}=\sigma_{32}\Delta N,
\end{equation}
where $\Delta$N is the inverse population density defined by Eq. (2).
The stimulated emission cross section $\sigma_{32}$ is given by
\begin{equation}
\sigma_{32}=\frac{\lambda_{32}^2}{2\pi}\frac{A_{32}}{2\pi\Delta\nu_D},
\end{equation}
where $\Delta\nu_D$  is  the Doppler width of the 3$\rightarrow$2
 transition.  At  a temperature  of $T \approx $100-1000 K in the 
 relatively cold HI region,  $\Delta\nu_D\approx$ (300-1000) MHz,
i.e., $\sigma_{32}=(0.6-2)\cdot 10^{-13}$ cm$^2$.    Thus, the
amplification coefficient may be estimated by the expression
\begin{equation}
\alpha_{32}=\sigma_{32}(W^{14}_{\mathrm {exc}}\tau_3)\beta N_{\mathrm {Fe}},
\end{equation}
where $\beta$ is  the fraction of the Fe$^+$ ions in state {\it 1} relative
to all Fe$^+$ ions (all iron atoms in the HI region of the blob
are supposed to be ionized).

      The  fraction $\beta$ is governed by the excitation  rate  of
level  {\it 1} and its radiative lifetime $\tau_1$. The excitation rates
for collisional population of level {\it 1} (i.e. by recombination
of  Fe  III  and  electron  collisions)  are  negligible  in
comparison with the radiative decay rate 1/$\tau_1$ because of the
low  electron density ($n_e\cong 10^4-10^5$ cm$^{-3}$) in the  HI  region.
The  electrons are generated in the photoionization  process
of  iron  as  well  as other elements having  an  ionization
potential  of  I.P$<$13.6  eV.  However,  the  most  important
channel  for  populating  the  metastable  state  {\it 1}  is  the
radiative decay of the high Fe II states selectively excited
by  Ly$\alpha$  or other intense lines. These excitation channels can
provide  for  an  excitation rate $>1/\tau_1\approx$ 1s$^{-1}$
  and  hence  sustain  the
relative  population  of state {\it 1} at a level  of  $\beta\approx10^{-2}$.
  This would  correspond to an approximately equal distribution  of
the Fe$^+$ ions among the 60 metastable states, including state
{\it 1}.  Leaving the calculation of a more accurate value of 
$\beta$ for future modeling, we adopt here a qualitative 
estimate of $\beta\approx10^{-2}$.   
An  evidence  for  a  substantial  population  in   the
metastable states is provided by the observation of  strong,
forbidden lines, e.g. from state {\it 1}, in the optical region of
$HST$ spectra of the blob (Zethson, 2001).

With these assumptions, we can use Eqs. (1) and (6) to
estimate the amplification coefficient for the 3$\rightarrow$2
 transition to $\alpha_{32}\approx(3\cdot10^{-18} - 10^{-16})N_{\mathrm {Fe}}$
 (cm$^{-1}$), where $N_{\mathrm {Fe^+}}\approx 10^{-4}\cdot N_{\mathrm H}$.  
 To create a cold HI region in the blob the hydrogen density  
 $N_{\mathrm H}$ must exceed a critical value, the so-called critical density
$N_{\mathrm {cr}}\approx  10^8$ cm$^{-3}$ 
(Johansson \& Letokhov 2001a,b). Thus, for 
$\alpha_{32}\geq (3\cdot10^{-14} - 10^{-12}$) cm$^{-1}$ and 
a blob diameter of $D\approx 10^{15}$ cm, which
can be regarded as the size $L$ of the amplifying region,
$\alpha L\approx (30 - 100)$   at an effective blob temperature
 of Ly$\alpha$  of T$_{\alpha}\approx (12-18)\cdot 10^3$ K.  The 
 amplification coefficient $\alpha_{32}$ is, according
 to Eq.7, sensitive to the spectral width of the amplified line.
 From this point of view the cold HI region is more suitable as a
 lasing medium than the hotter HII region, causing a higher Doppler
 width in the 3$\rightarrow$2 transition.
  
Let us emphasize that the amplification coefficient 
$\alpha_{32}$ according  to Eqs. (7)-(8) does not depend 
strongly  on  the
Einstein  coefficient  $A_{32}$, since $A_{32}\sim 1/\tau_3$  as  long  as  the
branching  fraction for the 3$\rightarrow$2 transition is dominating  in
the decay of state {\it 3}. The ($\alpha L$)-values obtained correspond  to
fairly   high   values   of   the   linear   (non-saturated)
amplification  coefficient $K$= exp($\alpha L$).  In  the  saturation   regime,
however,  the  intensity of the weak line $\lambda_{32}$  (in  photons/
cm$^2\cdot $s) approaches that of the strong line $\lambda_{43}$. 
Thereafter, the
amplification regime becomes saturated. This means that  the
rate of stimulated transition approaches the pumping rate of
level  {\it 3}  provided  by the spontaneous  4$\rightarrow$3 decay.
 Under such conditions the intensities of  both  these
lines grow in proportion to the propagation length $L$.

\section{ A radiative millisecond ``bottle neck'' and strong
 radiative cycling.}
                              
Many  of  the  radiative decay schemes of Ly$\alpha$   pumped  Fe  II
levels  contain  three  spontaneous transitions  before  the
electron ends up in a low-lying metastable state. In some of
these  schemes  there is a high probability that  the  final
state of the electron is the same as the initial state, i.e.
there  is  a high branching fraction of the final transition
connecting  states {\it 2} and {\it 1}. This is the case illustrated  in
Fig. 5a.
%
\\
\\
\noindent  \small
Fig. 6. A  partial energy level diagram (with  parameter
      values)  of  Fe II showing a closed loop  with  strong
      radiative  cycling.  The cycle  includes  a  selective
      photo-excitation of the highest level {\it 4}  by  intense
      Ly$\alpha$    radiation  and  a  millisecond  pseudo-metastable
      ``bottle-neck'',    causing    population     inversion.
      (Lifetimes  and  A-values are  from  Bergeson  et  al.
      (1996); Kurucz (2002); Raassen (2002)).
\normalsize
\\
\\
        First   we   discuss  the  closed  cycle  containing
Ly$\alpha$ absorption, spontaneous decay with emission of the bright UV
$\lambda\lambda$2507/2509 lines, stimulated emission at
1.68 and 1.74 $\mu$m, and a  fast  UV transition to the initial 
state. The diagram  in
Fig.  6 includes a few of the known energy levels of Fe  II,
which together with the inserted transitions and atomic data
are  relevant  in  the present study of a  strong  radiative
cycle.   The  broad  (a  few  hundreds  cm$^{-1}$) Ly$\alpha$
  line  almost
coincides in wavelength with two Fe II transitions from  the
low  level  a$^4$D$_{7/2}$ (level {\it 1} in Fig. 6) to the  close  levels
($^5$D)5p $^6$F$^o_{9/2}$ and (b$^3$F)4p $^4$G$^o_{9/2}$ 
(marked {\it 4} in Fig. 6), the frequency
difference  (detuning) being $\Delta\nu$=-160 cm$^{-1}$ and 
-185  cm$^{-1}$,  respectively.
The  detuning  is compensated by the Doppler  shift  or  the
broadening  of Ly$\alpha$ (see Sec. 8). The line at 2509.1  \AA\
  to  state  {\it 3}
(c$^4$F$_{7/2}$)  is the main decay channel from 
(b$^3$F)4p $^4$G$^o_{9/2}$ (state {\it 4})
with a branching fraction of about $\gamma \approx$ 0.9. Due to 
the small energy
difference there is an $ALM$ between the two energy levels  in
state {\it 4},  ($^5$D)5p $^6$F$^o_{9/2}$ and (b$^3$F)4p $^4$G$^o_{9/2}$,
 resulting in similar  decay
schemes (Johansson, 1978). As discussed above level {\it 3}  is  a
$PM$  state, with a lifetime of 1.5 ms, but in contrast  to  a
pure,  metastable  state  it can decay  slowly  by  electric
dipole radiation to the short-lived states {\it 2} 
(z$^4$D$^o_{7/2}$ and z$^4$F$^o_{9/2}$). To
get  strong radiative cycling it is important that  state  {\it 2}
decays by returning a large fraction of the Fe$^+$ ions to  the
initial  state  {\it 1},  which  is metastable  with  a  radiative
lifetime  of  about one second. Assuming the  blob  conditions
discussed in the previous section ($N_H \approx 10^9-10^{10}$ 
cm$^{-3}$) the time
scale for collisions is much longer than the time scale  for
radiative  decay  of  state {\it 3}. Under   these  conditions  the
closed  radiative cycle $1\rightarrow4\rightarrow3\rightarrow2\rightarrow1$
 has a ``bottle neck'' in the  $3\rightarrow 2$ transition.

       The  linear  amplification  coefficient  for  the  
$3\rightarrow 2$ transitions at 1.74 $\mu$m and 1.68 $\mu$m 
(Fig. 9) is defined  for $W^{14}_{\mathrm {exc}} << \tau_3^{-1}$
(a  rather  moderate  requirement)  by  Eqs.  (4)  and  (8),
rewritten as:
\begin{equation}
\alpha_{32} = \left[ \frac{\lambda_{32}^2}{2\pi} 
\frac{A_{32}}{2\pi\Delta\nu} \right] \tau_3 (\beta N_{\mathrm{Fe}}).
\end{equation}
From Eq. (9) we get that the amplification coefficient $\alpha_{32}$
(cm$^{-1}$)  for the  scheme  in Fig. 6 is largely determined 
by  the  factor($A_{32}\tau_3$). Its maximum value is 
$A_{32}\tau_3$=1 in the ideal case,  i.e.
when  there  is  only one radiative channel  of  spontaneous
decay  from state {\it 3}. In the real case (see Fig. 6) discussed
here  the  factor $A_{32}\tau_3 \approx$ 0.15. Rough estimates of 
$\alpha_{32}L \geq 10$,  where
$L\approx 10^{15}$  cm  is  the length of the amplifying medium,  can  be
obtained for $T_{\alpha} >10000$ K and $N_{\mathrm H} \simeq 10^8$ cm$^{-3}$
 or $T_{\alpha} >8000$ K and $N_{\mathrm H} \simeq 10^9$ cm$^{-3}$ .

      For an ensemble of atoms with sufficient  density
and  size  the  channel $3\rightarrow 2$ will open for a  fast  stimulated
decay,  and  the  duration of the whole radiative  cycle  is
determined  by  the  rate of the slowest excitation  channel
 rather  than  by  the $W^{14}_{\mathrm {exc}}$ rate of the slowest spontaneous  decay
$A_{32}$. For a high brightness temperature of Ly$\alpha$  in the range
$T_{\alpha}\approx$ 15000 K  the
excitation  rate is $W^{14}_{\mathrm {exc}} \approx 5\cdot 10^3$ s$^{-1}$
giving a duration  of  the  total
radiative cycle in Fe II of $\tau_{\mathrm {cycle}}\approx 2\cdot 
10^{-4}$ cycle. Hence,  the  Fe
ions  can  undergo  a  maximum  number  of  cycles  
($\tau_1/\tau_{\mathrm {cycle}} \approx 3.5\cdot 10^3$)
during the lifetime of the initial state ($\tau_1$).

      A  large  number  of  radiative  cycles  containing  a
stimulated channel provide a large intensity enhancement  of
the $\lambda\lambda$2507/2509 fluorescence lines ($4\rightarrow 3$)
 due to the suppressed
accumulation  of  Fe$^+$  ions  in  the  $PM$  state  {\it 3}  and  the
corresponding  depletion  of the initial  state  {\it 1}.  If  the
density  $N_{\mathrm {Fe}}$ and the size $L$ are not sufficiently  large  for
the  ensemble of atoms the amplification will be  small  and
the  stimulated channel will not operate. In such a case the
radiative  ``bottle neck'' limits the rate  of  the  radiative
cycling in Fe II to the time $\tau_1$=1.5 ms and the intensities of
the two UV lines $\lambda_{43}$ become normal.

      High-resolution  spectra of the  blobs  outside $\eta$ Car,
spatially resolved from the central star, show the anomalous
brightness of the two $\lambda\lambda$2507/2509 lines. Further observational
evidence in the blob spectrum for the radiative cycle is the
presence  of  one  ``bottle neck'' line at 8674.7  \AA\  (Zethson
2001),  which is an intercombination line with an  extremely
low  transition probability $A_{32} \approx$  13 s$^{-1}$ (Raassen  2002). The
``bottle neck'' transitions around 1.7 $\mu$m indicated in  Fig.  6
have  been observed in ground-based spectra of $\eta$ Car  (Hamann
et  al.,  1994). The 3$\rightarrow$2 transition from c$^4$F$_{7/2}$
 at $\lambda$=1.74 $\mu$m
and  1.68  $\mu$m appear relatively stronger than lines from  the
other  fine  structure levels of c$^4$F. Since the ground-based
spectra  of $\eta$ Car  contain integrated light  from  a  larger
region than the  Weigelt blobs discussed here the absolute
intensities cannot be compared with the intensities measured
in the $HST$ spectra.

      Moreover, according to the observations made in  (Gull
et al., 2001), the UV lines of the 2$\rightarrow$1 transition at 2740 and
2756  \AA\  have an integrated intensity comparable with that  of
the  bright lines of the 4$\rightarrow$1 transition. The width  of  these
spectral  lines is great because of the substantial  optical
density in the 4$\rightarrow$1 transition and the corresponding resonance
transfer  radiation broadening. Resonance radiation trapping
increases  the effective lifetime of level {\it 2}, but  still  it
remains much shorter than the long lifetime of level {\it 3}  and
does not prevent the formation of an inverted population  in
the 3$\rightarrow$2 transition.

\section{Radiative  microsecond ``bottle-neck'' and weak  radiative
cycling.}

In  some  of the schemes of Ly$\alpha$  pumped Fe II  levels
there  is  a  low probability that the final  state  of  the
electron is the same as the initial state, i.e. there  is  a
low  branching  fraction of the final transition  connecting
states  {\it 2} and {\it 1}. Such a case is illustrated in Fig. 5b.  The
primary  spontaneous decay of state 2 is to  the  metastable
state  {\it 1a} at 1 eV, and a weaker branch populates the initial
state  ({\it 1b} at about 3 eV) of the closed radiative cycle.  In
the case studied here state {\it 1a} is the same as state {\it 1} in the
millisecond ``bottle-neck'' discussed in Sec. 5, which  gave  a
strong radiative cycling. The low branching fraction of 0.1\%
for the 2$\rightarrow$1b transition at 5116 \AA\ produces a closed cycle but
a weak radiative cycling. Some feeding of the initial level,
state  {\it 1b},  is provided by the radiative cycle discussed  in
Sec.  5,  which means that there is only need for an initial
population  in state {\it 1a}   to start the pumping processes  of
both the ms- and $\mu$s- ``bottle-neck'' cycles.
%
\\
\\   
\noindent \small
Fig. 7. Diagram   showing  the  creation  of   population
     inversion  in the $3\rightarrow 2$ transition of Fe II because  of  a
     microsecond ``bottle-neck''. Since only a small  fraction
     of  the  Ly$\alpha$   pumped Fe II ions return  to  the  initial
     state  1b  the  closed loop 1b$\rightarrow4\rightarrow3\rightarrow2\rightarrow$1b  
generates  a  weak radiative cycling.
    \label{Fig. 7}
\normalsize
\\
\\
         A  more detailed scheme of the microsecond ``bottle-
neck''  with relevant atomic data is given in Fig. 7. The  Ly$\alpha$ 
radiation  populates a high-lying state of Fe  II  from  the
metastable   state  a$^4$G$_{11/2}$.  This  excitation   scheme   is
operative in various objects, verified by the observation of
the spontaneous radiative decay in the primary cascade as  a
fluorescence  line at 1869 \AA\ (Johansson \& Jordan,  1984)  as
well  as  the  secondary cascade at 9997  \AA,  seen  in  many
objects  as  a  prominent feature.  Based  on  the  extended
laboratory  analysis of Fe II (Johansson 1978), the  9997  \AA\
line  was first identified (Johansson, 1977) in the near-IR-
spectrum   of  $\eta$ Car  (Thackeray,  1969).  A   plausible
explanation   to   its  appearance  in   spectra   of   some
astrophysical objects was ascribed to the excitation  scheme
shown in Fig. 7 (Johansson, 1990).

      According  to theoretical calculations (Kurucz  2003),
the  main  radiative  decay from state  {\it 4},  
3d$^5$(a$^2$F)4s4p($^3$P)$^4$G$_{11/2}^o$ (hereafter sp$^4$G),
 occurs to the fine-structure levels
of  b$^4$G with branching fractions of 0.83 to J=11/2 and  0.04
to  J=9/2.  The most natural representative of  state  {\it 3}  in
Figs. 5b and 7 is thus b$^4$G$_{11/2}$ with a radiative lifetime  of
11 $\mu$s. Hence, it has a relatively slow spontaneous decay,
$A_{32}=8\cdot 10^4$ s$^{-1}$, to level {\it 2} (z$^4$F$^{o}_{9/2}$), 
which has a much shorter lifetime ($\approx$ 3 ns).
%
\noindent \small
\\
\\
Fig. 8. Comparison of HST/STIS spectra of the central star
      (flux  scale on right axis) and Weigelt blob  B  (flux
      scale   on   left  axis)  in $\eta$ Car  showing  intensity
      enhanced blob lines at 9913 and 9997 \AA\ (see text).
\normalsize
\\
\\
     Using  $HST/STIS$,  Gull  et  al.  (2001)  have  recorded
spectra  of  blob  B  in $\eta$ Car at high angular  and  spectral
resolution.  The spectrum contains the intense 9997  \AA\  line
(see Fig. 8), which can be excited by Ly$\alpha$  radiation according
to the excitation and decay scheme shown in Fig. 7. The 9997
\AA\  line  is  thus a transition from a microsecond $PM$  state,
which belongs to a different configuration (3d$^5$4s$^2$) than the
millisecond $PM$ state in Sec. 5. Since the lower state of the
9997  \AA\  transition is short-lived (a few  ns)  an  inverted
population  is built up, and the gain is determined  by  the
rate  of  photoexcitation provided by the Ly$\alpha$  radiation.  The
high intensity is naturally explained by stimulated emission
of  radiation,  and  the amplification coefficient  of  this
transition  is about one order of magnitude higher  than  in
the  case  of  the ms ``bottle-neck'' because  of  the  factor
($A_{32}\tau_3$) in Eq. (9).

\section{Combination  of strong and weak radiative  cycles  in  a
millisecond ``bottle neck''.}

     The  existence  of a great number of metastable  states
and  the  high density of high-lying energy levels  in  FeII
causes several coincidences between absorption lines and  Ly$\alpha$ 
and  thereby the generation of inverted population. In  this
section  we give examples of  cascade schemes from the  same
millisecond  ``bottle  neck'' but with both  strong  and  weak
radiative cycles.

     In  Figure  9 we present two transition schemes,  which
provide  for  the amplification of the extremely  weak  FeII
lines  at 9391.5 \AA,  9617.6 \AA\ and 9913.0 \AA\ and some  other  lines
around  2 $\mu$m. The difference between the two groups is only  the  $PM$
states, which have different J-values but belong to the same
$LS$  term.  The  two schemes represent a combination  of  the
previous  ones discussed in Secs. 6 and 7 due  to  the  fact
that  there  are  two  sets  of possible  ``bottle  neck''  
3$\rightarrow$2
transitions.  In one set state {\it 2} consists of  sextet  levels
explaining the extremely low transition probability  $A_{32}$  in
the  intercombination lines at $\lambda\lambda$9391/9617 in Fig. 9a  and
$\lambda$9913 in Fig. 9b. Since the sextet levels decay mainly to the ground  term in Fe II and only in a very slow decay  to  the
initial level, the closed 
1$\rightarrow4\rightarrow3\rightarrow2\rightarrow$1 radiative 
cycle becomes very
weak.   However,  like the case studied in  Sec.  6  the  $PM$
states  can also decay to quartet levels in state  {\it 2},  which
combine  strongly with the initial state. The  corresponding
$PM$ transitions occur in the near-IR region around 2 $\mu$m.
%
\\
\\
\noindent \small
 Fig. 9. Partial energy level diagrams of Fe II showing two
      examples   of  combined  strong  and  weak   radiative
      cycling  due to alternative decay routes for  the  two
      ``pseudo-metastable'' states c$^4$P$_{3/2}$ (Fig 9a) and
      c$^4$P$_{1/2}$
      (Fig.   9b).   The   ``bottle-neck''  effect   generates
      inverted  population and lasing in the 3$\rightarrow$2  transitions
      to   sextet  and  quartet  levels,  which  have  quite
      different  branching  fractions (BF)  to  the  initial
      state 1.
\normalsize
\\
\\
     The  amplification coefficient for the 2 $\mu$m lines  should
be  of the same order of magnitude as for the 1.7 $\mu$m lines in
Fig. 6 in Sec 5, whereas it is reduced for the lines in  the
range  9300-9900 \AA\ due to the smaller $A$ value. The reduction
of  the  Einstein  coefficients $A_{32}$ has,  in  principle,  no
effect  on the amplification coefficient, provided that  the
upper  level  {\it 3}  decays radiatively only to  one  low  laser
level.  However, this is not the case here, and  the  faster
decay  to  the  quartet levels in a  time  of $\tau_3\approx A_{32}^{-1}$
  reduces  the
amplification  coefficient (6) for the  transitions  to  the
sextet levels by a factor of $A_{32}\tau_3 \approx 10^{-2}$. 
But, the great gain margin is
sufficient to ensure $\alpha L \geq10$, even for the weak lines, and all three
lines  mentioned  above are observed and identified  in  the
blob spectrum of $\eta$ Car (Zethson 2001). Two of them, the laser
lines  at  9617 \AA\ and 9913 \AA, are shown in Fig. 10 as  intense
features  in  the  blob spectrum but  not  observed  in  the
stellar  spectrum. Moreover, the two lines at 9617.6  \AA\  and
9391.5  \AA\  should be very weak in spontaneous  emission  and
have  an intensity ratio of 2.5, determined by the branching
ratio (the ratio of the $A$-values). These lines, however, are
observed  (Gull  et al. 2001) as intense  lines  with  equal
intensities. This is naturally explained by their  formation
through  a  stimulated mechanism, which is  operative  under
saturated excitation conditions with a common pumping source
and  a common upper level. The ``peculiar'' intensity ratio 
observed for the $\lambda\lambda$9391/9617 lines forms perhaps 
the strongest argument for lasing (Hillier, private communication). 
The laser lines around 2 $\mu$m in  the
strong  radiative cycle of Fig. 9a,b appear, like  the  1.7 $\mu$m
lines  in  Fig.  6,  as  prominent features  in  the  ground
spectrum  of $\eta$ Car (Hamann et al 1994). The absence  of  the
laser lines in the stellar spectrum (see Fig. 10) is due  to
collisional  relaxation  of the $PM$  states  in  the  stellar
atmosphere/wind    implying    a    Boltzmann     population
distribution.
%
\\
\\
\noindent \small
Fig. 10. Comparison of HST/STIS spectra of the central star
      (flux  scale  on  right axis) and the Weigelt  blob  B
      (flux  scale on left axis) in $\eta$ Car with the two lasing
      lines $\lambda\lambda$9617,9913 generated in the level  schemes  in
      Fig. 9 and discussed in Sec 7.
\normalsize
\\
\\
\section{Sources  of  Ly$\alpha$  Radiation  Pumping Fe  II  in  the
Weigelt blobs}

The blobs in $\eta$ Car are embedded in the photospheric radiation
and  the  wind  from the central star. Let us  speculate  about
possible sources for Ly$\alpha$  radiation capable of pumping  Fe  II
within the blobs.
     Firstly, the Ly$\alpha$  radiation emitted by the photosphere of
$\eta$ Car  can excite Fe$^+$ ions in blobs located at distances of
$R_b \approx (300-1000)\cdot r_s$,
where $r_s \approx 10^{13}$  cm  is  the  stellar radius. But,  
even  at  such  a
relatively close location of the blob the dilution factor  
$\Omega = (r_s/2R_b)^2 = 10^{-6}-10^{-5}$,
which   substantially  weakens  the  intensity  of  the   Ly$\alpha$ 
radiation  reaching  the  blob and  makes  it  incapable  of
producing   the  bright  2507/2509  \AA\  lines  (Johansson   \&
Letokhov, 2001).

     The stellar wind of $\eta$ Car is an important characteristic
of  the star's surroundings, the mass loss rate of the  star
being  substantial (Davidson \& Humphreys, 1997). Spherical
(Hillier et al., 2001) as well as aspherical (Smith et  al.,
2003)  models  of the stellar wind of $\eta$ Car with  a  terminal
velocity of $v_1\simeq 500-1000$ km/s have been considered. 
In accordance with models
by  Lamers  \&  Cassinelli (1999) of  the  stellar  wind  its
velocity  is  almost close to the terminal value  everywhere
between $\eta$ Car and the GC. The Lyman continuum radiation, Ly$_c$ ,
from  the  star photoionizes the stellar wind and after  the
subsequent recombination a wide Ly$\alpha$  profile is produced. 
 Another potential source of ionizing radiation is a binary companion,
but in the discussion below about Ly$\alpha$ formation inside the blob 
it is not important to distinguish between the individual components 
in a binary system.
It is  possible to consider three cases of irradiation  of  the
blob  by  radiation from the stellar wind depending  on  the
size  of  its Str\"{o}mgren radius, $\tilde{R}_{\mathrm {Str}}$
  relative to the distance  $R_b$
between the blob and the star as illustrated in Fig. 11a, b,
c.
%
\\
\\
\noindent \small
Fig. 11. A  schematic model of a  Weigelt blob,  located at a 
     distance R$_\mathrm{b}$  from  the  hot,
     luminous   star $\eta$ Car  and  subject  to   the   stellar
     radiation  and  wind.  (a) A local Str\"{o}mgren  boundary,
     R$_{\mathrm{str}}$,  between  the  H II and H I  regions  is  located
     inside  the  blob.  Stellar  radiation  in  the   Lyman
     continuum (Ly$_c$ , h$\nu>$h$\nu_c=13.6$ eV) photo-ionizes hydrogen  only
     in  the front part of the blob. Radiation in the  range
     7.6  eV  $<h\nu<13.6$ eV passes through the H II region and  photo-
     ionizes iron atoms, which have a  density  of $N_{\mathrm{Fe}}\simeq
10^{-4}N_\mathrm {H}$  in
     the  H  I region. Intense Ly$\alpha$   radiation from the  H  II
     region  excites  Fe II resonantly in  the  H  I  region
     (lasing   volume).  The  Str\"{o}mgren  boundary  for   the
     stellar wind $\tilde{R}_{\mathrm {str}}>>R_\mathrm{b}$ (b) the stellar wind 
is optically  thick
     for  Ly$_c$   and the blob is a one-component H  I  medium.
     Only  remote,  blue-  shifted Ly$\alpha$    radiation  from  the
     stellar wind can irradiate the HI region containing  Fe
     II;  (c) the Str\"{o}mgren boundary in the wind lies behind
     the   blob   ($\tilde{R}_{\mathrm {str}}\approx R_\mathrm{b}$),  providing  
      intense,  red-shifted   Ly$\alpha$ photons for efficient excitation of Fe II.
\normalsize
\\
\\
     1)$\tilde{R}_{\mathrm {Str}}>>R_b$   (Fig. 11a). In this 
case the Ly$_c$  radiation reaches
the blob without significant attenuation and can photoionize
the  front part of it. If the hydrogen density in  the  blob
exceeds the critical density $N_{\mathrm {cr}} \simeq 10^8$
cm$^{-3}$ (Klimov, Johansson \& Letokhov,
2002)  the  Str\"{o}mgren boundary of the  blob,  defined  by  ,
intersects the blob. The front part of the blob produces  an
intense Ly$\alpha$  radiation with $T_{alpha} >$10000 K. Thus, the  dilution  of
the  photospheric Ly$\alpha$ radiation falling on to the blob is to  a
great degree compensated by spectral compression of absorbed
broadband radiation to a $10^3-10^4$ times more narrow Ly$\alpha$  profile
(Johansson  \&  Letokhov, 2004). Under these  conditions  the
blob is very different from a typical planetary nebula since
the  dilution  factor $\Omega =(r_s/2L)^2$ is {\em many orders of  
magnitude  greater} than for a planetary nebula 
($\Omega \simeq 10^{-15} - 10^{-12}$).

     2) $\tilde{R}_{\mathrm {Str}}<<R_b$  (Fig.  11b). 
In this case the Lyman  continuum  is
blocked out for the blob. This means that the entire  volume
of  space  between $\eta$ Car and the blob is  an  ionized  H  II
region,  with  the Str\"{o}mgren boundary for the  stellar  wind
being  located between the star and the blob. Since hydrogen
is  not fully ionized the blob does not have two regions  as
in  the  previous  case, where the  HII  and  HI  zones  are
adjacent  to each other (Fig. 11a). A single-component  blob
(a  HI region) is probably too cold to emit the bright UV Fe
II  lines. It will therefore make sense to assume  that  the
optical  density of the Lyman continuum $\tau (\lambda_c)\leq 1$
in  the  region between $\eta$ Car and the blob.

     3) $\tilde{R}_{\mathrm {Str}}\geq R_b + D$  (Fig.  11c).  
In this case the Str\"{o}mgren  boundary
outside the blob is located slightly behind it. This can  be
considered as an optimal situation for pumping of Fe  II  in
the blob for two reasons. Firstly, the  Ly$\alpha$ photons produced  in
this torus-like zone beyond the blob are red-shifted and 
$\Delta \lambda_w = \lambda_0(v_t/c)cos\Theta$, where 
$\Delta \lambda_w$ = the wavelength shift, $\lambda_0$=  the  rest
wavelength, $v_t$= the terminal stellar wind velocity  and $\Theta$  is
the  viewing angle as indicated in Fig. 11c. The red-shifted
and  broadband Ly$\alpha$  radiation irradiates the H I region of the
blob  without any need for diffusion Doppler broadening.  It
can  penetrate  the  whole  H I region,  since  its  optical
density $\tau (\lambda_0 + \Delta \lambda$) in the far wing of 
Ly$\alpha$  is not abnormally  high,
where $\Delta \lambda$=2.4 \AA\ is the shift of the Fe II absorption  line
relative to the center of Ly$\alpha$.

     Secondly,  the dilution factor for the Lyman  continuum
irradiating the stellar wind (see geometry in Fig. 11c) will
largely  be  compensated for by the conversion of  Ly$_c$   into
narrower Ly$\alpha$  radiation in the stellar wind in the vicinity  of
the  blob.  The optimal case is when the size of $\tilde{R}_{\mathrm {Str}}$  coincides
with the distance to the irradiated part of the HI region in
the  blob, which will require a significant density  of  the
stellar wind near the blob. The critical hydrogen density in
the stellar wind, $\tilde{N} _{cr}$,  that can provide 
$\tilde{R}_{\mathrm {Str}}\simeq R_b$, is connected with the
critical density of the blob itself, $N_{cr}$, required to locate
the  Str\"{o}mgren  boundary inside the blob, by the  approximate
expression:
\begin{equation}
\tilde{N}_{cr}/N_{cr} \simeq (D/R_b)^{1/2}.
\end{equation}
For  the adopted values of the distance to the star
$R_b \simeq 300-1000r_s \simeq 10^{16}-3\cdot 10^{16}$ cm and the
size  of  the  blob, $D \simeq 10^{15}$ cm, $\tilde{N} _{cr}$ 
is some tenths of $N_{cr} \simeq 10^8$ cm$^{-3}$, i.e a  relatively
high  density  of the stellar wind close to the  blob.  This
very  approximate estimate of $\tilde{N} _{cr}$ in the vicinity of  the  blob
(0.15\arcsec  from $\eta$ Car)  agrees with other estimates,  based  on
observed  and modelled spectra of the stellar wind  (Hillier
2001,  2003). The evaluation of Eq. (10) agrees  with  direct
measurements of the size and shape of the stellar  wind  and
the  blobs in $\eta$ Car (van Boekel et al., 2003). Of course, all
these qualitative speculations should be subject of detailed
modelling  of  the stellar wind with the blob  included,  but
that is out of the scope of the present paper.

     In  Secs.  4  and  5  above we have used  the  effective
spectral  temperature of Ly$\alpha$ ($T_{\alpha} >$10000 K), in the treatment  of
the  pumping  of  absorption lines to achieve  a  population
inversion  in  some Fe II transitions from $PM$  states.  This
requirement is valid for any possible source of  intense  Ly$\alpha$ 
radiation. The consideration of possible Ly$\alpha$ pumping from the 
stellar wind is still preliminary and qualitative, and it needs
a future detailed modelling. 

\section{On the spectral brightness and spectral width of 
astronomical lasers}

We have emphasized that the radiation intensity (photons$\cdot$
s$^{-1}$$\cdot$cm$^{-2}$) of an astrophysical laser (APL) line cannot 
be higher than the intensity of the pumping line. In principle the 
spectral brightness (photons$\cdot$s$^{-1}\cdot$cm$^{-2}\cdot$Hz$^{-1}\cdot$
sr$^{-1}$) of the APL line can exceed that of the pumping line 
because of two effects: 1) the narrowing of the APL line by resonant 
amplification of the radiation (both the spontaneous radiation itself 
and some external radiation), and 2) an anisotropic type of amplification 
(for example in an elongated amplifying region). Since these effects 
are interrelated, they can be considered simultaneously.

In the regime of linear amplification of spontaneous emission, the line width 
decreases because of an exponential amplification at the line center. This is 
described by a simple relation (Casperson \& Yariv 1972; Allen \& Peters 1972): 
\begin{equation} 
\Delta\nu = \frac {\Delta\nu_D}{\sqrt{1 + \alpha L}}, 
\end{equation}
 where $\alpha$ is the gain coefficient (see Sec. 4) per unit 
length. The narrowing occurs until the amplified line, which is inhomogeneously 
Doppler-broadened, gets saturated and its peak "flattens out". This effect was 
considered for the APL-saturated amplifier, and it has been shown (Litvak 1970) 
that the narrowing of the spectrum lines ceases at saturation. The line broadens 
again to the width of the gain line, which has been confirmed by observational 
data for astrophysical masers (Reid \&Moran 1981; Elitzur 1992). 

It would be very interesting to measure the sub-Doppler line width of the 9997 
\AA\ line (as well as other laser lines mentioned above) using a spectral 
resolution R $> 10^6$ and an angular resolution better than 0.1 arcsec. This 
could, for example, be done with spatially separate telescopes using a 
heterodyne correlation analysis of the Brown-Twiss effect (Johnson et al, 1974). 
The spectral brightness of APL radiation can grow as a result of 
the narrowing of the spectrum lines. In order to observe this effect, one will 
need a detection technique that is capable of very high spectral resolution. 
Such methods are known in radio-frequency spectroscopy, but are as yet 
unavailable in the optical region. However, such measurements in the optical
region using Brown-Twiss-Townes heterodyne interferometry are in principle 
quite feasible (Lavrinovich \& Letokhov 1976; Letokhov 1996). An analysis of
performing such a direct proof of the laser effect in the Weigelt blobs will
be discussed elsewhere (Johansson \& Letokhov, in preparation). 

\section{ Conclusion: On the difference of the evidence of astrophysical 
lasers and masers.}

     Space lasers with optical pumping operate according  to
a  scheme  similar  to  the  one  suggested  for  the  first
laboratory laser 45 years ago by A. Schawlow and  C.  Townes
(1958).  We  believe that laser amplification and stimulated
emission  of  radiation  is a fairly common  and  widespread
phenomenon,  at  least  for  gaseous  condensations  in  the
vicinity  of bright stars. This is due to the occurrence  of
two types of processes (fast radiative and slow collisional)
in  the  very  rarefied gaseous condensations,  whereby  the
population of energy levels in atoms (ions) can relax. These
relaxation processes occur on highly different time  scales,
radiative  relaxation  operating on a  wide  time  scale  of
$10^{-9}-10^{-3}$  s   (sometimes  even  within  $10^{-3}$-1 s),   and
collisional relaxation, on a time scale of over seconds  (at
gas  densities  $<10^{10}$ cm$^{-3}$). In the case  of  photoselective
excitation of some high-lying electronic levels of  an  atom
or  ion  with  a  complex energy-level structure,  radiative
relaxation  can  take  place as a  consequence  of  downward
transitions with spontaneous emission of radiation,  in  the
course of which there develops an inverse population of some
pair  (pairs) of levels. If the size of a gas cloud is large
enough,   large  amplification  on  the  inverted-population
transition   automatically   switches   on   the   radiative
relaxation   channel,  which  leads  to  faster   stimulated
transitions until collisional relaxation becomes  important.
Thus, the laser action is an intrinsic characteristic of the
radiative  cooling  of  gas  clouds  near  bright  stars  by
stimulated  emission  for inverted  transitions  along  with
spontaneous emission for normal transitions.

      However,  the  detection of the laser effect  is  less
evident  in the optical region of the spectrum than  in  the
microwave  region, where it gives rise to  strong  radiation
lines   of   exceptionally   high   brightness   temperature
($ 10^{10}-10^{15}$  K (Elitzur, 1992)). In the optical  region,  the
laser effect raises the intensity of weak forbidden lines up
to the intensity of those strong allowed lines providing the
selective  photoexcitation  of the  upper  levels.  This  is
explained by the huge difference ( $\approx 10^{15}-10^{18}$ times)  between
the   spontaneous  emission  rates  and  inverse  population
production  mechanisms in the two wavelength  regions.  This
follows from simple qualitative considerations of the steady-
state saturation regime of the isotropic maser/laser action.

      The  stimulated emission of radiation in a space maser
occurs  as  a  result of pumping of the upper  maser  level,
which  is  not associated with any spontaneous  emission  in
radiative  microwave transitions. It may  therefore  have  a
decay  rate much in excess of the spontaneous emission rate,
which lies in the region $A_{\mathrm {mn}} \simeq 10^{-9}-10^{-7}$ s$^{-1}$.
 As a result, the intensity of the
stimulated radiation can be many orders of magnitude  higher
than  the intensity of the spontaneous radiation and  it  is
only limited by the pumping rate. This is exactly the reason
why  the  brightness  temperature of maser  microwave  lines
reaches  as  high a value as $ 10^{10}-10^{15}$ K. The  intensity  of
maser  lines  is not borrowed from other microwave  spectral
lines, which are very weak, but from other pumping sources.

      In  the  optical region of the spectrum, the  rate  of
allowed  spontaneous  transitions is  high  
($A_{\mathrm {mn}} \simeq 10^{8}-10^{9}$ s$^{-1}$),  and  it  is
precisely  spontaneous  transitions  that  provide  for  the
pumping  of the upper level in an optical space laser  at  a
sufficiently  high rate to exceed the rates  of  collisional
pumping  mechanisms. This is especially true  for  the  case
considered  here  where the space laser is being  indirectly
pumped  by  HLy$\alpha$  in the vicinity of $\eta$ Car, one of 
the most luminous stars of our Galaxy. Therefore, the  intensity  of
the stimulated radiation in the optical region generated  by
the  occurrence  of an inversion population and  significant
amplification  cannot  in  the steady-state  regime  to  any
substantial  extent  exceed the  intensity  of  the  pumping
spectral  lines, formed by spontaneous emission  in  allowed
radiative  transitions of atoms or ions. This fact  presents
difficulties   in  detecting  laser  action   by   a   large
enhancement of the intensity of the radiation, but it should
manifest  itself in comparatively moderate  changes  of  the
branching  ratios of spectral lines, having a common  source
of pumping.

     One exception is the case of quantum transitions having
a  relatively low spontaneous emission probability  
($A_{\mathrm {mn}} \simeq$ a few - 10$^{5}$ s$^{-1}$),  and
consequently the spontaneous radiation lines are weak.  Once
an  inverted  level  population  has  developed  in  such  a
transition  with a significant amplification in  a  properly
sized  cloud,  a stimulated emission channel opens  up.  The
stimulated  transition rate cannot be much higher  than  the
spontaneous  emission rate, which is limited by the  pumping
rate  of the upper laser level. Thus, the intensity  of  the
laser  line  should become comparable with the intensity  of
the lines, which are due to allowed spontaneous emission  of
radiation and resulting from the optical pumping (direct  or
indirect) of the upper level. This takes place only  in  the
inverted  population  volume  with  size $L >> 1/\alpha$
,  where $\alpha$  is  the
amplification coefficient per unit length. In such a case, a
spectral  line,  expected  to be  weak,  must  appear  as  a
spectral  line of normal intensity, typical for  an  allowed
transition.  This is exactly what we have found  to  be  the
case  with several spectral lines of Fe II pumped indirectly
by  the  intense  H Ly$\alpha$  radiation (1215 \AA)  in  the
Weigelt blobs of $\eta$ Car.

       In  conclusion,  let  us  emphasize  that  stimulated
emission  at  optical and near-infrared wavelengths  forming
non-collisional,  radiative cycling of atomic  particles  in
astrophysical  media  gives  a natural  explanation  to  the
anomalous behaviour of spectral lines of two types:  i)  {\em the
transformation of a very weak line from a $PM$  state  have  a
low  transition probability (line 3$\rightarrow$2 in Figs. 1a, 6,  9)
into  a strong line with an intensity comparable to that  of
an  allowed  transition}  and ii)  the  transformation  of  a
normally intense line into an anomalously bright line  (line
4$\rightarrow$3 in  Fig.  10) due to radiative cycling,  which  is  
{\em only possible if stimulated emission drives the weakest  link  in
the cycle.}

Before astrophysical spectra at very high resolution can be observed 
these conclusive facts constitute, at present, the only
proofs of the Fe II astrophysical laser in the Weigelt blobs of $\eta$ Carinae. 

\begin{acknowledgements}  
{We thank Dr. John Hillier for sending us data on  the
density  of the stellar wind, and later for providing very useful 
referee comments on the manuscript. V.S.L. acknowledges  financial
support  through  grants (S.J.)  from  the  Royal  Swedish
Academy  of  Sciences and the Wenner-Gren Foundations, 
as well as Lund Observatory for hospitality  and
the  Russian  Foundation for Basic Research  (grant No. 103-02-
16377). The research project is supported by a grant  (S.J.)
from the Swedish National Space Board.}
\end{acknowledgements}

\end{document}